\documentclass[%
 reprint,
 amsmath,amssymb,
 aps,
]{revtex4-2}

\usepackage{graphicx}% Include figure files
\usepackage{dcolumn}% Align table columns on decimal point
\usepackage{bm}% bold math
\usepackage{hyperref}% add hypertext capabilities
\begin{document}

\preprint{APS/123-QED}

\title{Observation of Alfv\'en Wave Reflection in the Solar Chromosphere: Ponderomotive Force and First Ionization Potential Effect}% Force line breaks with \\
%\thanks{Distribution Statement A: Approved for Public Release; Distribution is Unlimited.}%

\author{Mariarita Murabito$^1$, Marco Stangalini$^2$, J. Martin Laming$^3$, Deborah Baker$^4$, Andy S. H. To$^4$, David M. Long$^{5,6}$, David H. Brooks$^7$, Shahin Jafarzadeh$^{8,9}$, David B. Jess$^{5,10}$, Gherardo Valori$^7$}

\affiliation{$^1$INAF Istituto Nazionale di Astrofisica, Osservatorio Astronomico di Capodimonte, 80131 Napoli, Italy}%}\footnote{currently at INAF Istituto Nazionale di Astrofisica, Osservatorio Astronomico di Roma, 00078, Monteporzio Catone, Space Science Data Center (SSDC) - Agenzia Spaziale Italiana, via del Politecnico, s.n.c., I-00133, Roma, Italy}
 \email{mariarita.murabito@inaf.it}
%
%\author{Marco Stangalini}%
\affiliation{$^2$ASI -- Italian Space Agency, Via del Politecnico, s.n.c 00133 – Rome, Italy}
%
%\author{J. Martin Laming}%
\affiliation{$^3$Space Science Division, Code 7684, Naval Research Laboratory, Washington DC 20375, USA}
%
%\author{Deborah Baker}
\affiliation{$^4$University College London, Mullard Space Science Laboratory, Holmbury St. Mary, Dorking, Surrey, RH5 6NT, UK}
%
%\author{David M.~Long}
\affiliation{$^5$Astrophysics Research Centre, School of Mathematics and Physics, Queen’s University Belfast, University Road, Belfast, BT7 1NN, Northern Ireland, UK}
\affiliation{$^6$School of Physical Sciences, Dublin City University, Glasnevin Campus, Dublin, D09 V209, Ireland}
%
%\author{David H. Brooks}
\affiliation{$^7$Department of Physics \& Astronomy, George Mason University, 4400 University Drive, Fairfax, VA 22030, USA}
%
%\author{Andy S. H. To}
%\affiliation{University College London, Mullard Space Science Laboratory, Holmbury St. Mary, Dorking, Surrey, RH5 6NT, UK}
\affiliation{$^8$Max Planck Institute for Solar System Research, Justus-von-Liebig-Weg 3, 37077 G\"{o}ttingen, Germany}

\affiliation{$^9$Niels Bohr International Academy, Niels Bohr Institute, Blegdamsvej 17, DK-2100 Copenhagen, Denmark}

\affiliation{$^{10}$Department of Physics and Astronomy, California State University Northridge, Northridge, CA 91330, USA}

\date{\today}% It is always \today, today,

\begin{abstract}
We investigate the propagation of Alfv\'en waves in the solar chromosphere, distinguishing between upward and downward propagating waves. We find clear evidence for the  reflection of waves in the chromosphere and differences in propagation between cases with waves interpreted to be resonant or nonresonant with the overlying coronal structures. This establishes the wave connection to coronal element abundance anomalies through the action of the wave ponderomotive force on the chromospheric plasma, which interacts with chromospheric ions but not neutrals, thereby providing a novel mechanism of ion-neutral separation. This is seen as a ``First Ionization Potential Effect'' when this plasma is lifted into the corona, with implications elsewhere on the Sun for the origin of the slow speed solar wind and its elemental composition.
\begin{description}
\item[Usage]
\item[Structure]

\end{description}
\end{abstract}

\maketitle

%\tableofcontents

%\section{\label{sec:level1}Introduction}
The first hint of varying elemental composition in the solar atmosphere came in 1963 \citep{pottasch1963}. Taken seriously in the 1980's following influential reviews \citep{meyer1985a,meyer1985b}, the phenomenon was systematized as an enhancement in coronal abundances by about a factor of 3 in elements such as Fe, Mg, and Si with first ionization potential (FIP) less than about 10 eV
\citep{Feldman2000,Feldman2003}. High-FIP elements, those with FIP greater than 10 eV (e.g. O, Ne, He, and S), are relatively unaffected. This fractionation has become known as the ``FIP Effect''.

In the mid 1990s, imaging time series of solar atmospheric dynamics revealed significantly larger-than-anticipated wave motions in the corona, e.g. \citep{EIT1999,nakariakov1999}. In the absence of an explanation for the FIP Effect in terms of ``conventional'' processes such as the various forms of diffusion or gravitational settling, this last observation suggested a role for magnetohydrodynamic (MHD) waves interacting with plasma ions (but not with neutrals) via the ponderomotive force \citep{Laming2004}. The basic mechanism involved a coronal loop that acts as a resonant cavity for MHD waves, while wave reflection/refraction in the high density gradient in the solar chromosphere can separate the ions (low-FIP elements) from the neutrals (high-FIP elements). This new feature in solar physics is an analogy in MHD of the trapping of particles in optical physics by the refraction of laser beams \citep{ashkin1970,ashkin1986}. 

Observations of elemental abundances now encompass a wide variety of coronal regions and solar wind regimes \citep{Warren2014,Baker2013,Ko2016,Feldman1998,Baker2018,vonsteiger2000,Zurbuchen2012}.
%, as well as the coronae of late-type stars of varying spectral types \citep{Laming1996,Drake1997,Laming1999,Wood2018,Seli2022}. Many of these 
These different fractionations can be reproduced by the ponderomotive force model, where simple changes to the magnetic geometry in the chromosphere-corona region introduce modifications to the MHD wave fields, which in turn alter the fractionation in realistic ways. For example, the reduced FIP fractionation in the fast solar wind from polar coronal holes compared to the slow wind from more equatorial solar regions is easily reproduced by modern models \citep{Laming2015}. It has also been well established that turbulence in the fast solar wind observed close to Earth orbit \citep{bruno2013,Ko2018} is more Alfv\'enic and less balanced (i.e. more outgoing than ingoing waves) than in the slow wind. %Closer to the Sun similar results hold, but with unbalanced Alfv\'enic waves also being found in slow wind periods \citep{damicis2021,zhao2022}. 
%Along with the abundance modification, these results too should be a relic of waves in the fractionation region.
Along with the abundance modification, {if caused by waves}, these {properties} too should be seeded by {the same} waves connected to the fractionation
region.

\begin{figure*}
\includegraphics[scale=0.28, trim=350 0 400 0]{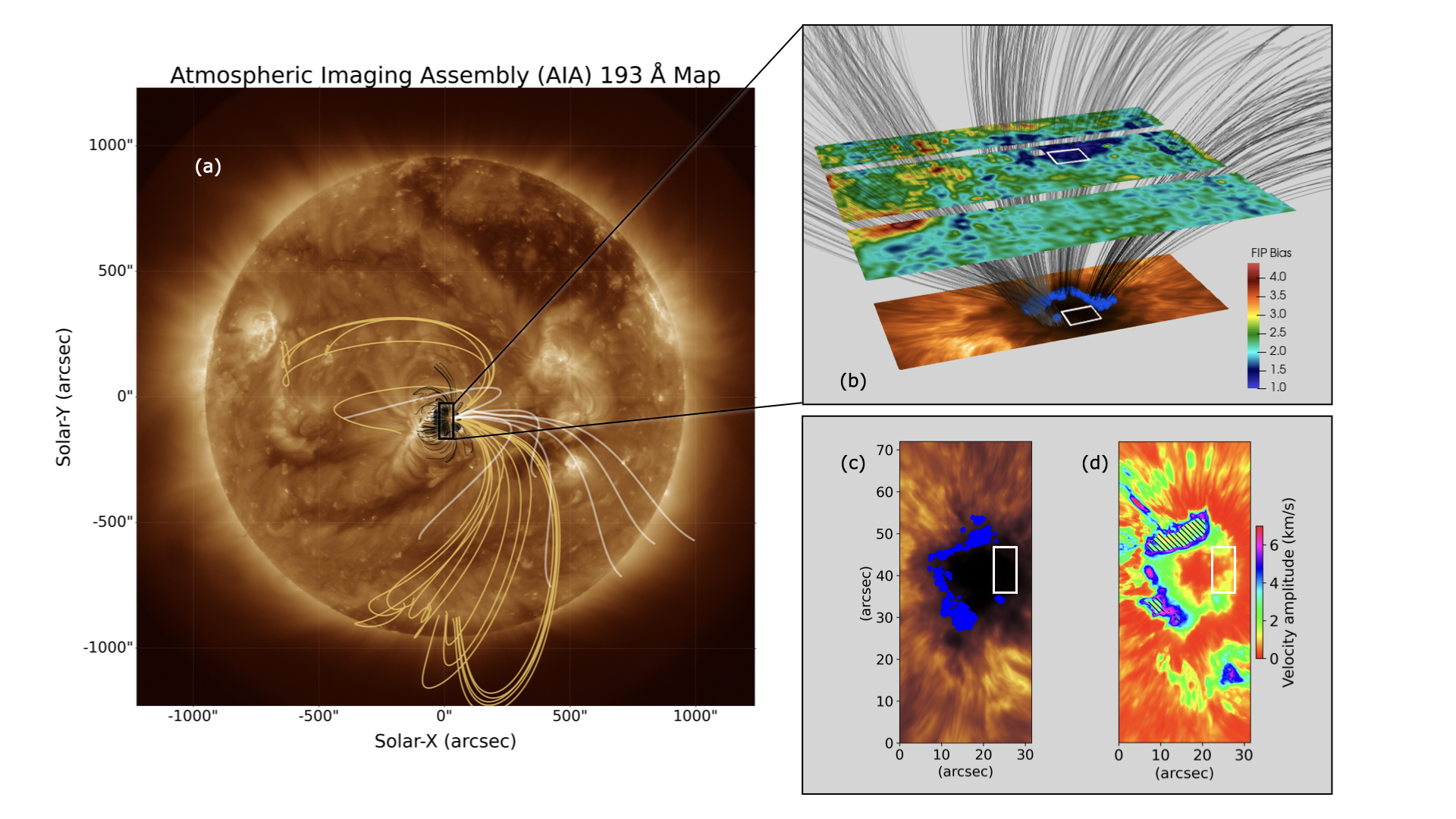}
\caption{Panel (a): Full-disk SDO/AIA 193~\AA\ coronal map with selected field lines from the PFSS extrapolation based on a SDO/HMI synoptic magnetogram, showing three populations: small loops (black), big loops (yellow), and open field lines (white). Panel (b):  3D overview of the region of interest linking the chromospheric disturbances (``Blue Dots'') and locations of strong coronal FIP-bias through {a linear force free} extrapolation, {rotated by nearly 90$^{\circ}$ for better visualization.} Panel (c): Chromospheric DST/IBIS Ca~{\sc{II}} 8542.2 nm line-core intensity image showing Blue Dots and Box A locations. Panel (d): Chromospheric velocity amplitude of Alfv\'en(ic) waves. The hatched area shows where the Ca~{\sc{II}} Stokes $I$ intensity profile goes into emission. Adapted from \citep{Baker2021}. \label{fig1}
}
\end{figure*}

Thanks to a combination of high-resolution ground-based spectropolarimetric observations of the solar chromosphere and space observations of the solar corona, regions of strong coronal FIP bias were linked for the first time to the existence of chromospheric magnetic waves in a sunspot \citep{Stangalini2021,Baker2021}. %In particular, using magnetic field extrapolation, the authors were able to connect regions (called `blue dots', BD hereafter) with enhanced fractionation to chromospheric oscillations detected using the line-of-sight component of the magnetic field \citep{Murabito2021}. 
So far, the existence of MHD wave reflection/refraction in the chromosphere where FIP fractionation happens has only been theorized, and their cross-helicity (``balance'') only measured {\it in situ} in the solar wind. %The focus of this study is on the chromospheric reflection mechanisms, and we report in particular the first observational evidence that the reflection of waves occurs in the chromosphere at locations magnetically linked to high FIP bias in the corona in a manner consistent with the theory of element abundance modification by the ponderomotive force. 
In this work we study Alfv\'en wave balance and reflection in the chromosphere, and reinforce our ideas about FIP fractionation.

%The dataset used in this work has been presented in a series of papers \citep[e.g.,][]{Stangalini2021,Baker2021,Murabito2021} and has been the main focus of other studies \citep[see, e.g.,][]{Stangalini2018,Murabito2019,Houston2020} due to the high quality of the data and the large-scale nature of the observed sunspot, which was the leading spot of AR~12546. 
The target of this study was acquired with the Interferometric BIdimensional Spectrometer \citep[IBIS;][]{Cavallini2006} instrument at the Dunn Solar Telescope (DST) on 2016 May 20 for more than two hours of observations. The same portion of this active region (i.e., the large sunspot) was also observed using the EUV Imaging Spectrometer \citep[EIS;][]{Culhane2007} on board the \emph{Hinode} satellite \citep{Kosugi2007}. These specific data were used to make a spatially resolved map of coronal composition. In detail, the ratio of the low-FIP Si~{\sc{x}} 258.375\AA~to the high-FIP S~{\sc{x}} 264.223\AA~diagnostic was used to compute the FIP bias map \citep{Baker2021}. The ground-based observations are stored in the IBIS-A archive \citep{Ermolli2022} and can be accessed at \href{http://ibis.oa-roma.inaf.it/IBISA}{http://ibis.oa-roma.inaf.it/IBISA}. 

\begin{figure*}
\includegraphics[scale=0.9]{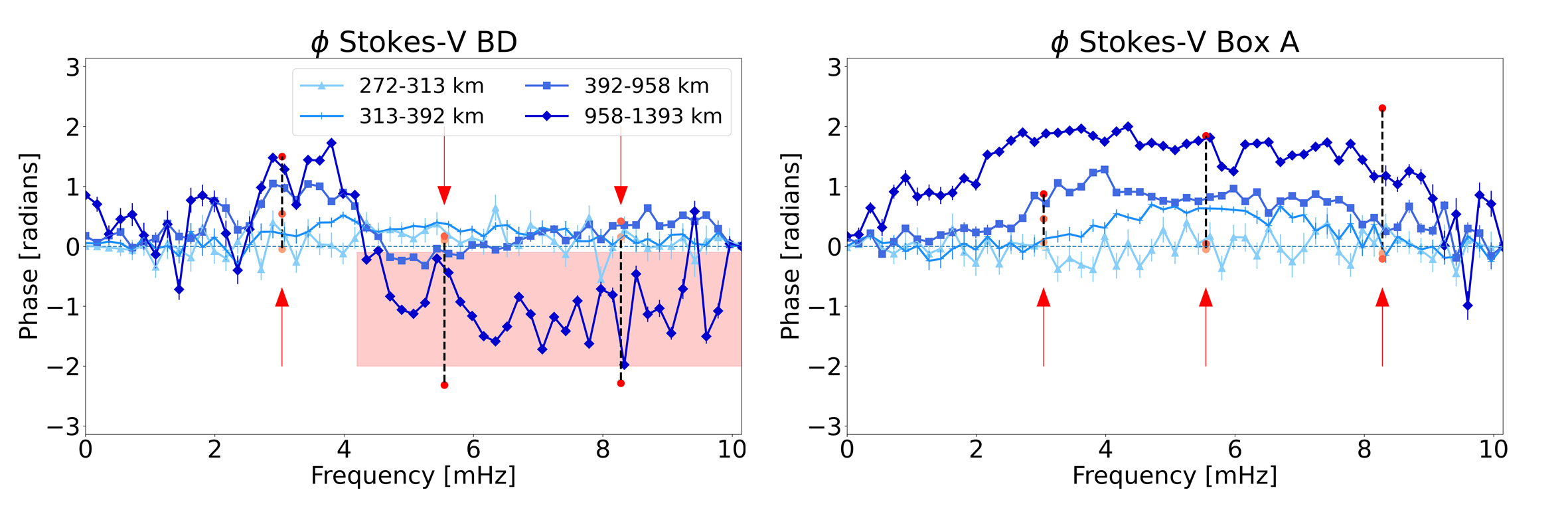}

\caption{\label{fig2} Phase lag diagrams between 300 km and 1300 km in the solar atmosphere from the Ca~{\sc{ii}} 854.2 nm Stokes $V$ (the first eleven spectral positions) in the two regions BD (left) and Box~A (right) as identified in Figure~\ref{fig1}c. All plots are obtained by considering only those phase values for which the coherence is larger than 90\% in each frequency bin. Modelled phase shifts are shown from lightest to darkest red for both BD and Box~A. These use slightly different height intervals to the observations (see text), but the qualitative agreement is clear. The shaded area indicates the change in the propagation of the waves.}
\end{figure*}

An overview of the observations described above is presented in Figure~\ref{fig1}. Panel (a) shows a full disk coronal image with a superimposed potential
field source surface (PFSS) extrapolation. {Within the limits of the approximation \citep{riley06,linker17,loeschl24} this should give a good representation of the large scale solar magnetic field structure}. Panel (b), adapted from \citep{Baker2021}, shows a coronal FIP bias map. The western side of the sunspot shows a FIP bias of between 1.5 and 2. {Here, a linear force free} extrapolation shows that this region (``Box A'') is magnetically connected to another active region or has open field lines, the yellow and white lines in panel (a), respectively. By contrast higher FIP bias values of 3-4 are found on the eastern/southern sides of the spot in the active region core loops (black field lines), connected with  locations of enhanced chromospheric wave power, shown as ``Blue Dots'' (BDs). These are overplotted on the ground-based high-resolution Ca~{\sc{ii}} 854.2 nm line-core intensity image in panel (c). Panel (d) shows the velocity amplitude of the Alfv\'enic waves.
The strong Alfv\'enic perturbations in the chromosphere \citep{Stangalini2021} are magnetically linked to coronal loops containing highly fractionated plasma \citep{Baker2021} as expected {if waves cause the fractionation}.

The ground-based observations consist of full Stokes ($I$, $Q$, $U$, and $V$ measurements) scans of the Ca~{\sc{ii}} 854.2~nm lines, acquired along with 21 spectral points with a cadence of 48~s. Taking into account the Ca~{\sc{ii}} formation height, i.e., from line wing to line core, the 21 spectral points sample the solar atmosphere spanning from the mid-photosphere ($\sim$300~km formation height, on average) to the mid-chromosphere \citep[$\sim$1300~km formation height;][]{Murabito2019}. These multi-height observations allow us to investigate wave activity through the solar atmosphere \citep{Jess2023}. Using a Fourier phase-lag technique (see the Appendix~\citep{suppl}) it is possible, by employing a stringent coherence confidence level, to retrieve important wave characteristics as they propagate between various layers of the solar atmosphere. 

In order to perform this task, we calculate the phase angles and associated coherence levels over the whole frequency spectrum between pairs of Ca~{\sc{ii}} 854.2~nm Stokes $V$ (i.e., linked to the longitudinal magnetic field) through consecutive spectral points. From the first 9 measured spectral points, we are able to explore the coupling between the upper photosphere and lower chromosphere \citep{delacruz2013,Bjorgen2018}.
To have reliable phase measurements, we only include those that have a coherence level greater than 90\%, which ensures the accuracy and statistical significance of our identified waves \citep{Jess2023}. %We follow a similar approach used in previous studies, i.e., where we compare the wave properties over the BDs region with an area located on the opposite side where the FIP bias is lower than 3. The latter is marked in Figure~\ref{fig1}c and will be referred to as `Box~A'. 
We are helped by the unusually strong magnetic field in the sunspot, taken to be 1500 G in the penumbra chromosphere \citep{Murabito2019,Stangalini2022}.

Figure~\ref{fig2} shows the phase-lag diagrams for these two regions (i.e., BD and Box~A on the left and right side, respectively) using the Stokes $V$ time series. In each panel, the curves are plotted from the lightest to darkest colors, associated with the lowest to highest solar atmospheric heights (i.e., 300~km above the solar surface up to 1300~km in the solar chromosphere). The geometric-height approximations, corresponding to the spectral points, were derived using the RH code \citep{Uitenbroek2001}, assuming the FALC atmospheric model \citep{suppl,Fontenla1993}, to be consistent with the NLTE inversion results used here and in previous works. {While the FALC atmospheric model assumes a quiet Sun region, the difference between the chromosperic structure in quiet Sun and sunspot regions do not contribute to significant differences in the derived heights of each of the spectral points.} Waves travel upwards/downwards when positive/negative phases are detected, respectively. All the magnetic waves (considering the Stokes $V$ measurements) detected inside Box~A (right panel of Figure~{\ref{fig2}}) travel upwards from the photosphere into the chromosphere. This is {\emph{not}} the case within the BD region, where the low frequency waves show positive phases and move upwards, while negative phases, indicating downward motion, for frequencies $>4$~mHz are detected at upper-atmospheric heights (958 -- 1393~km; darker blue lines in the left panel of Figure~{\ref{fig2}}). This indicates a source of waves higher up in the atmosphere, or possibly in the opposite chromosphere (see the red dots or the shaded area in the left panel of Figure~{\ref{fig2}}).

\begin{figure*}
\includegraphics[scale=1, trim= 0 0 0 0]{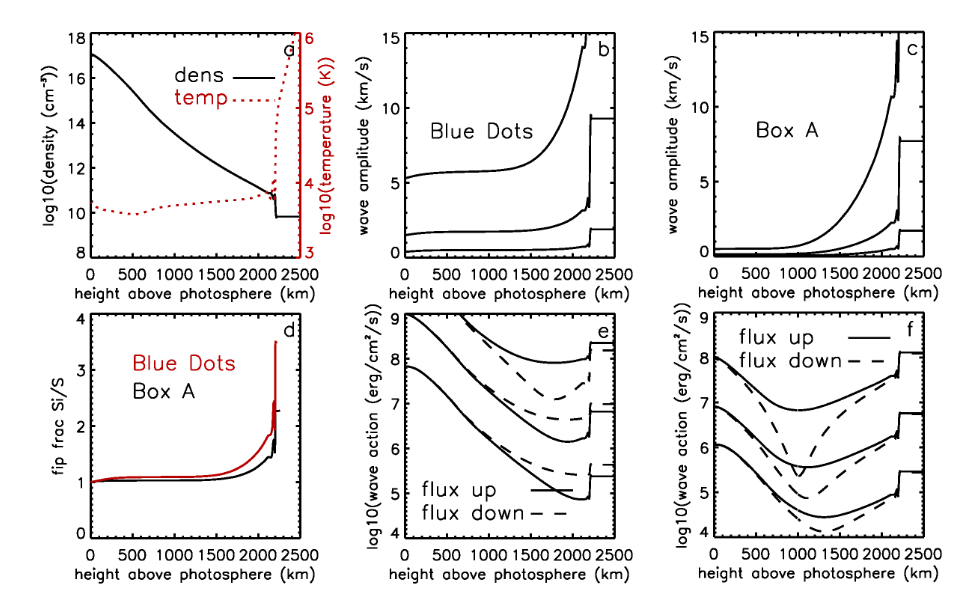}
%\vspace{-1.0in}
\caption{\label{fig3} Wave propagation models used to interpret observations. Panel (a):FALC chromosphere density and temperature profiles. Panels (b) and (e): Wave amplitudes and energy fluxes for the Blue Dots (BDs) region. In the order of decreasing wave amplitude or flux, the wave frequencies are 3.04, 5.55, and 8.3~mHz. Panels (c) and (f): Same plots (as in (b) and (e)) for waves in Box~A, where these frequencies represent the loop fundamental, as well as 1st and 2nd harmonics. Panel (d): Si/S fractionation resulting from the ponderomotive force in each region. {The discontinuities in all panels at 2250 km arise from the steep chromosphere-corona density gradient.}}
\end{figure*}

%\section{Discussion and Conclusion}
In the most developed models of the FIP fractionation, \citep{Laming2004,Laming2015,Laming2017,Laming2019}, ion-neutral separation arises in the chromosphere as a result of the time independent part of the ponderomotive force associated with the passage of Alfv\'en and/or fast mode waves. This force is dictated by the wave field profile as a result of wave propagation in the structured atmosphere, with reflection due to density and magnetic-field gradients, as well as nonlocal effects arising as the waves interact with coronal structures (most easily characterized in terms of resonance). A wave is considered resonant with a coronal loop if the wave travel-time from one loop footpoint (a location of strong wave reflection) to the other is an integral number of wave half-periods. 

The PFSS extrapolation reveals three 
populations of magnetic structures connecting to the umbra. There are small core loops on 
the east side, depicted in black, connecting two footpoints within the same active 
region. These have lengths, $L$, of order 
of 10--100~Mm and magnetic fields of 50--500~G (comparable with that reported by 
\citep{Brooks2021} using the observations of Fe~{\sc{x}}). Together with 
coronal densities, calculated using the observed EIS Fe~{\sc{xiii}} 
\citep{Baker2021}, around $10^9$ cm$^{-3}$, these yield resonant frequencies, 
$\nu =V_A/L$, where $V_A$ is the Alfv\'en speed, in the range of 0.1--1~Hz. 
Longer loops, shown in yellow, connect the western side of the sunspot to other active regions and are about 1000--2000~Mm long, with magnetic fields $\sim 10$ G, that, with coronal densities in the $10^7-10^8$~cm$^{-3}$ range give loop resonant frequencies of 1--10~mHz, i.e., the range observed in the present work. 
Finally open fields, identified by white curves, are less relevant to this study, and may actually be part of the population of longer loops that are sufficiently long to leave the computational domain.

We construct models of the wave propagation presented in Figure \ref{fig3}. Waves are excited at about 1000 km altitude in one chromosphere, and the transport equations \citep{suppl} are integrated to the other footpoint to give the behavior there as displayed. Panel (a) shows the density and temperature profiles from the chromospheric model FALC \citep{Fontenla1993}, the same model used in the NLTE spectropolarimetric inversions reported in \citep{Murabito2019}. Panel (d) illustrates the FIP fractionation for Si/S for both the BD and Box~A. Panels (b) and (e) respectively display the wave amplitudes and energy fluxes for the BD region. In the order of decreasing amplitude or flux, the wave frequencies are 3.04, 5.55, and 8.3~mHz. Panels (c) and (f) show the same plots as in (b) and (e), but for waves in Box~A, where their associated frequencies represent the loop fundamental, as well as the 1st and 2nd harmonics. At chromospheric heights of 1500~km, the directionality of the waves shown in Figure~\ref{fig3} panel (e) and (f) agrees with that observed in Figure~\ref{fig2} for both the BD and Box~A regions. Furthermore Figure~\ref{fig3} panel (d) provides the observed FIP fractionation for each case. {The ponderomotive acceleration has been calculated from the Alfv\'en wave amplitudes with parametric generation of sound waves taken into account, following \citep{Laming2015}}.
At lower atmospheric heights, around 500--1200~km, the directionality is preserved better in the 3.04 mHz waves, and is reduced for the higher frequencies, as also in the observations. At even lower geometric heights, all waves have approximately equal upward and downward components, in agreement with the observations. Phase shifts of the magnetic field perturbations in the models are plotted with red symbols on Figure \ref{fig2}, for the height intervals 180-210, 210-252, 252-276, 276-1385 km for Box A and 180-210, 210-252, 252-560, 560-1385 km for the BDs. It is worth noting that {for a given matching the frequency range between observations and our models, the resulting height intervals} are slightly different. This reflects the difference in reflection properties between the model and real sunspot chromospheres. {However these heights are all well below the chromospheric
altitude where the fractionation occurs in our models, and so no serious error is expected to result. Fiducial representations of the solar atmosphere at the time of the observations are not possible since we do not have corresponding observations of the opposite loop footpoints nor of the loop coronal sections. Wave damping is neglected, but both the parametric generation of sound wave and the turbulent cascade of counter propagating Alfv\'en wave should be expected, especially in the chromospheric regions below 1000 km where the Alfv\'en speed decreases (approaching the sound speed) and the magnetic field lines become more curved, increasing the perpendicular component of the wavevector. 

As well as Alfv\'en wave propagation, the FIP fractionation depends on the ionization balance for various elements. In the cases of H and Ca, these are shown not to vary dramatically from their values in static chromospheric models \citep{Carlsson2002,Wedermeyer2011}. For fractionation at the top of the chromosphere, where the density gradients are steep and the background gas is ionized, no fractionation happens at all unless the element neutral fraction is less than about 0.5\% (the ratio of scattering rates of neutrals and ions with the background gas) \citep{Laming2019}. Ionization beyond the singly charged state makes no difference because the ponderomotive acceleration for low frequency shear or torsional Alfv\'en waves is the same for all charged particles.}

The resonant frequencies of the coronal structures in Box~A (i.e., 1--10~mHz) are in the range detectable with IBIS observations. We interpret these results in the right hand panels of Figure \ref{fig2} as the coronal structure picking out the resonant frequency from a continuum of chromospheric sound waves. 
%The ponderomotive acceleration and fractionation are therefore characteristics of resonant waves, in that they are strongly peaked at the steep density gradient at the top of the chromosphere where H is becoming ionized. 
Except for the swapping of downgoing and upgoing flux, the wave solutions in each chromosphere are identical for resonant waves.

The BD loop resonances at 0.1--1~Hz are much higher than the frequencies of the observed chromospheric waves, which are now nonresonant.  We assume the same wave frequencies as for Box~A, calculated in the same way, with about twice the amplitude as before. These waves are now off resonance, but reproduce the observed Si/S fractionation of about 3.5. The velocity amplitudes observed in Ca {\sc{II}} in Fig. \ref{fig1}b are now much higher than in Box A in Fig. \ref{fig1}c, because wave energy is not trapped in the coronal loop and drains out to the chromosphere.
As well as the FIP fractionations, the calculated wave reflection characteristics also match well with observations.

%Although the plasma fractionation in the solar atmosphere has been explored extensively, on different timescales (solar cycle, magnetic flux emergence and decay, and solar flares) and for the various solar regions (quiet-Sun, coronal-hole and active regions) \citep{Warren2014,Baker2013,Ko2016,Feldman1998,Baker2018}, combined multi-heights observations still present a challenge. The observations given in \citep{Stangalini2021,Baker2021,Murabito2021} represent a unique extant example. These 
Previous studies \citep{Stangalini2021,Baker2021,Murabito2021} highlighted the existence of magnetic oscillations as a driver for the coronal fractionation. This aspect has always been the key ingredient of the ponderomotive force introduced for separating ions from neutrals in the chromosphere. 
We go further in this work and infer that the magnetic oscillations, detected and linked with the high FIP bias, are directional in the chromosphere and corona as predicted by models of the fractionation. The resonant waves are more balanced in the coronal sections of the loop than are the nonresonant waves, suggesting that as well as the abundances, the different turbulence properties in various regimes of the solar wind might be {seeded by} their solar origins. We emphasise that the wave properties are dictated by conditions at each end of the coronal loop and the degree to which the wave is resonant with the loop, and so the fractionation is inherently nonlocal.

\begin{acknowledgements}
 MM is currently at INAF Istituto Nazionale di Astrofisica, Osservatorio Astronomico di Roma, 00078, Monteporzio Catone, Space Science Data Center (SSDC) - Agenzia Spaziale Italiana, via del Politecnico, s.n.c., I-00133, Roma, Italy.
 
 This work was supported by the Italian Space Agency (ASI) with the scientific contribution to Solar-C under contract to the co-financing National Institute for Astrophysics (INAF) Accordo ASI-INAF 2021-12-HH.0 ``Missione Solar-C EUVST – Supporto scientifico di Fase B/C/D''.
 MM acknowledge financial support from the ASI-INAF agreement n. 2022-14-HH.0.
 JML was supported by NASA HSR Grant NNH22OB102 and by Basic Research Funds of the Office of Naval Research. ASHT thanks the STFC for support via funding given in his PhD studentship. DML is grateful to the UK Science Technology and Facilities Council (STFC) for the award of an Ernest Rutherford Fellowship (ST/R003246/1). 
 DBJ acknowledges support from the UK Space Agency for a National Space Technology Programme (NSTP) Technology for Space Science award (SSc~009). DBJ is also grateful to the UK STFC for additional funding via the grant awards ST/T00021X/1 and ST/X000923/1. DBJ also wishes to thank The Leverhulme Trust for grant RPG-2019-371. 
 GV acknowledges funding by the Bundesministerium für Wirtschaft und Technologie through Deutsches Zentrum für Luft- und Raumfahrt e.V. (DLR), Grants No. 50 OT 1001/1201/1901 as well as 50 OT 0801/1003/1203/1703, and by the President of the Max Planck Society (MPG).
 Hinode is a Japanese mission developed and launched by ISAS/JAXA, collaborating with NAOJ as a domestic partner, and NASA and STFC (UK) as international partners. 
 Scientific operation of Hinode is performed by the Hinode science team organized at $ISAS/JAXA$. 
 This team mainly consists of scientists from institutes in the partner countries. 
 Support for the post-launch operation is provided by JAXA and NAOJ (Japan), STFC (UK), NASA, ESA, and NSC (Norway). 
 DB is funded under STFC consolidated grant number ST/S000240/1.
 The work of DHB was performed under contract to the Naval Research Laboratory and was funded by the NASA Hinode program.
 Finally, we wish to acknowledge scientific discussions with the Waves in the Lower Solar Atmosphere (WaLSA; \href{https://www.WaLSA.team}{https://www.WaLSA.team}) team, which has been supported by the Research Council of Norway (project no. 262622), The Royal Society \citep[award no. Hooke18b/SCTM;][]{2021RSPTA.37900169J}, and the International Space Science Institute (ISSI Team~502). 
\end{acknowledgements}

Author contributions -- Conceptualization: M.M., M.S., D.B.; Formal Analysis: M.M., J.M.L., A.S.H.To, S.J.; Investigation, Methodology, Project administration, Resources, Software, Visualization: M.S.; Writing-original draft: All; Writing-review and editing: All.

\end{document}